\documentclass[twocolumn,secnumroman,amssymb, amsmath, nofootinbib,
nobibnotes, aps, prb, showpacs]{revtex4}

\usepackage{graphicx}% Include figure files
\usepackage{dcolumn}% Align table columns on decimal point
\usepackage{amsmath}
\usepackage{here}% 
\begin{document}

\preprint{HEP/123-QED}

\title{Electronic states realized by the interplay \\ 
between Li diffusion and Co$^{\rm 3+}$$/$Co$^{\rm 4+}$ charge ordering in Li$_{\rm x}$CoO$_2$
} 

%\title{Enhanced resistivity after rapid cooling in Li$_{\rm x}$CoO$_2$:\\
%Interplay between Li diffusion and Co$^{\rm 3+}$$/$Co$^{\rm 4+}$ charge ordering} 

\author{Kiyotaka Miyoshi,$^1$ Kentaro Manami,$^1$ Ryo Sasai,$^1$ Shijo Nishigori,$^2$ and Jun Takeuchi$^1$}
\affiliation{%
$^1$Department of Material Science, Shimane University, Matsue 690-8504, Japan
}%
\affiliation{%
%\address{%
$^2$ Department of Materials Analysis, CIRS, Shimane University, Matsue 690-8504, Japan
}%

%\affiliation{%


%\affiliation{
%\affiliation{
%\affiliation{%


\date{\today}% It is always \today, today,

\begin{abstract}

Measurements of dc magnetization ($M$) and electrical resistivity ($\rho$) have been carried out as 
a function of temperature ($T$) for layered oxide 
Li$_{\rm x}$CoO$_2$ (0.51$\leq$$x$$\leq$1.0) using single crystal specimens. 
After slow cooling of the specimens down to 10 K, both of the $M$($T$) and $\rho$($T$) curves are found to 
exhibit a clear anomaly due to the occurrence of Co$^{\rm 3+}$$/$Co$^{\rm 4+}$ charge ordering (CO)
at $T_{\rm S}$$\sim$155 K for 0.6$\leq$$x$$\leq$0.98 (at $T_{\rm S}$$\sim$180$-$190 K for 0.5$\leq$$x$$\leq$0.55). 
After rapid cool of the specimens, 
additional anomalies are observed related to the onset of Li diffusion 
at $T_{\rm F1}$$\sim$370 K and$/$or $T_{\rm F2}$=120$-$130 K. 
Due to phase mixing with compositions of nearly LiCoO$_2$ and Li$_{\rm 2/3}$CoO$_2$, 
the specimens with 0.7$\lesssim$$x$$\lesssim$0.9 show anomalies both at $T_{\rm F1}$ and $T_{\rm F2}$. 
For 0.6$\lesssim$$x$$\lesssim$0.9, 
the resistivity measured after rapid cooling is found to 
be fairly larger than that measured after slow cooling below $T_{\rm S}$. 
The enhanced resistivity can be explained by the scenario that disordered Co$^{\rm 3+}$$/$Co$^{\rm 4+}$ arrangements, which have 
been observed and revealed to have an insulating electronic structure contrasting to the regular CO state in the previous 
scanning tunneling microscopy measurements [Phys. Rev. Lett. {\bf 111}, 126104 (2013)], are 
realized due to the formation of an amorphous-like structure of Li ions after rapid cooling via interlayer Coulomb coupling. 
An electronic phase diagram for 0.5$\leq$$x$$\leq$1.0 is proposed. 
%It is found that the specimens with 0.7$\lesssim$$x$$\lesssim$0.9 contains two hexagonal phases with compositions 
%nearly LiCoO$_2$ and Li$_{\rm 2/3}$CoO$_2$, where Li diffusion occurs above $T_{\rm F1}$ and $T_{\rm F2}$, respectively. 
%which is expected to  with 
%Two onsets are originating from 
%phase mixing in the specimens with 
%0.7$\lesssim$$x$$\lesssim$1.0 having compositions nearly LiCoO$_2$ ($T_{\rm F1}$) and Li$_{\rm 2/3}$CoO$_2$ ($T_{\rm F2}$). 

\end{abstract}

%\pacs{75.40.Cx, 71.27.+a, 75.30.Kz}% PACS, the Physics and Astronomy
\maketitle

%\tableofcontents

\section{\label{sec:level1}Introduction}

Li-ion batteries are now being widely used for mobile electronic devices and 
also electric vehicles, and have been the subject of intensive study 
to improve storage capacity and longevity.\cite{goodenough,nitta} 
One of the most common cathode materials used in commercial Li-ion batteries with 
high-energy density and a proper application voltage is layered oxide Li$_{\rm x}$CoO$_2$, 
in which both Li and Co atoms are octahedrally coordinated by oxygen atoms, forming 
a two-dimensional (2D) regular triangular lattice in each layer, and 
each layer is alternatively stacked along the c-axis. 
Much attention has been focused on the high mobility of Li ions in 
Li$_{\rm x}$CoO$_2$ for the application but various unconventional physical 
natures are expected as in the related material Na$_{\rm x}$CoO$_2$, which has a similar 
structure consisting of Na and CoO$_2$ layers. 
Indeed, Na$_{\rm x}$CoO$_2$ has been under intensive study for the past two decades due to 
the various intriguing electronic properties, 
such as a large thermoelectric power ($x$$\sim$0.7),\cite{terasaki} superconductivity ($T_{\rm c}$$\sim$5 K)
in the hydrated compound ($x$$\sim$0.35),\cite{takeda} an insulating ground state induced by 
charge ordering ($x$$\sim$0.5),\cite{foo,yokoi,gasparovic,ning} a mass-enhanced Fermi-liquid ground state analogous to 
LiV$_2$O$_4$\cite{miyoshi}, a spin-density-wave state,\cite{motohashi} in addition to 
the characteristic Na order,\cite{zand,chou1,weller,meng,roger,julien,morris,alloul,
mukhamedshin,schulze,galeski,medarde} which has been found to be linked with Co$^{\rm 3+}$$/$Co$^{\rm 4+}$ charge disproportionation,
\cite{meng,roger,julien,morris,alloul,mukhamedshin} magnetic ordering,\cite{schulze,galeski} 
and structural transitions.\cite{medarde} 

In both compounds, Co ions are in a mixed valence state consisting of Co$^{3+}$ ($t^{6}_{\rm 2g}$) and Co$^{4+}$ ($t^{5}_{\rm 2g}$) 
having spins of $S$=0 and $S$=1$/$2, respectively,\cite{ray} since the deintercalation of Li or Na ions from the mother compounds 
generates Co$^{4+}$ on the 2D triangular lattice of Co$^{3+}$. 
Thus, a hole doping in the $t_{\rm 2g}$ orbital can be made by the deintercalation in Li$_{\rm x}$CoO$_2$, 
where the electrical resistivity drastically decreases and a metallic behavior appears in the $\rho$($T$) curve 
with decreasing $x$ from 1.0 to 0.9.\cite{menetrier,kellerman,miyoshi10,motohashi11,lin} 
One of the most characteristic features of Li$_{\rm x}$CoO$_2$ is the Co$^{\rm 3+}$$/$Co$^{\rm 4+}$ charge ordering (CO), 
which has been inferred from sharp anomalies observed in the dc magnetization,\cite{kellerman,sugiyama05,mukai07,hertz,motohashi09} 
resistivity\cite{miyoshi10,motohashi11}, and specific heat\cite{miyoshi10} data at $T_{\rm S}$=150$-$170 K 
in wide range of $x$, and found to have a characteristic $\sqrt{3}$$\times$$\sqrt{3}$$R$30$^{\circ}$ 
arrangement of Co$^{\rm 4+}$ by recent scanning tunneling microscopy (STM) observations 
and density functional theory (DFT) calculations.\cite{iwaya} 

The ordering of Co$^{\rm 3+}$$/$Co$^{\rm 4+}$ can be strongly affected by the dynamics of Li ions via interlayer Coulomb coupling. 
In a previous study, we have reported that the dc magnetization ($M$) versus temperature ($T$) curve 
measured after rapid cooling of the specimens is 
different from that measured after slow cooling below $T_{\rm F}$$\sim$120 K for $x$=0.66,\cite{miyoshi10} 
suggesting that Li ions stop diffusing and become ordered below $T_{\rm F}$, 
since $T_{\rm F}$ is similar to $T_{\rm MN}$ and $T_{\rm d}^{\rm Li}$ determined by the measurements of 
NMR\cite{nakamura} and muon spin rotation ($\mu$SR)\cite{sugiyama09}, respectively, as the temperature below which Li ions start diffusing. 
The origin of the difference below $T_{\rm F}$ is attributed to the difference in the Co$^{\rm 3+}$$/$Co$^{\rm 4+}$ arrangement, which is expected to be 
disordered when the Li ions form an amorphous-like structure as in a glassy state 
after a rapid cooling down to a temperature 
far below $T_{\rm F}$. Indeed, it has been revealed in the previous STM observations at $T$$\sim$5 K 
for Li$_{0.66}$CoO$_2$ that 
there is some disordered area on the crystal surface 
where the CO state is destroyed and an insulating gap in the $dI/dV$ spectrum is detected. 
Thus, the cooling-rate-dependent Li ordering might play a decisive role 
in determining the Co$^{\rm 3+}$$/$Co$^{\rm 4+}$ arrangement. 

High-quality single-crystal specimens of Li$_{\rm x}$CoO$_2$ used in the previous study have enabled us to 
understand the electronic properties through microscopic measurements\cite{iwaya,ikedo,mizokawa,simonelli,okamoto}, but 
the interplay between Li diffusion and Co$^{\rm 3+}$$/$Co$^{\rm 4+}$ ordering, and the evolution of 
the charge ordering for the delithiation have remained unsolved. 
In the present paper, we have performed dc magnetization and electrical resistivity measurements for Li$_{\rm x}$CoO$_2$ 
using single-crystal specimens with a systematic change of Li content $x$ (0.5$\leq$$x$$\leq$1.0) to shed further light 
on the problems. It has been found that electrical resistivity for the specimens with 0.6$\lesssim$$x$$\lesssim$0.9 
measured after rapid cooling of the specimens becomes much larger than that 
measured after slow cooling below $T_{\rm S}$$\sim$155 K. The results are consistent with the above-mentioned scenario 
that the region where the CO state is destroyed, having an insulating electronic structure, appears due to 
the amorphous-like structures in the Li layers induced by the rapid cooling. We have proposed an electronic 
phase diagram for 0.5$\leq$$x$$\leq$1.0 based on the results of the measurements 
using 12 single-crystal specimens with different $x$. 
\section{Experiment}
Single crystal specimens of Li$_{\rm x}$CoO$_2$ (0.51$\leq$$x$$\leq$1.0) were obtained by 
chemically delithiating from LiCoO$_2$ single-crystals,  
as described in a previous report.\cite{miyoshi10} 
In the first step, single crystals of Na$_{0.75}$CoO$_2$ were grown in an optical floating-zone 
furnace in a similar manner as described in the literature.\cite{chou04} 
The obtained single crystal rods of Na$_{0.75}$CoO$_2$ were crushed into small pieces, and then cleaved 
into thin slices with a thickness of $\sim$0.2 mm. 
In the next step, to obtain LiCoO$_2$ single crystals by ion exchange reactions, 
the cleaved Na$_{0.75}$CoO$_2$ single crystals were embedded in Li$_2$CO$_3$ powder in an alumina boat, and 
heated 600 $^{\circ}{\rm C}$ for 24 h in air, and then repeatedly washed with acetonitrile 
to remove Li$_2$CO$_3$ and Na$_2$CO$_3$. After annealing these crystals 
at 900 $^{\circ}{\rm C}$ for 24 h in air, single crystal specimens of pristine LiCoO$_2$ were obtained, which have a 
plate-like shape with a typical dimension of 2$\times$2$\times$0.2 mm$^3$. 
In the final step, the delithiation from pristine LiCoO$_2$ crystals was carried out by chemically extracting lithium using 
NO$_2$BF$_4$ as the oxidizer. The reaction was carried out in an argon atmosphere by immersing LiCoO$_2$ crystals 
in an acetonitrile solution of NO$_2$BF$_4$ and heating at 50 $^{\circ}{\rm C}$ for 72 h 
in a pressure vessel with a Teflon liner, and finally washed to remove LiBF$_4$ by acetonitrile. 
The Li content $x$ (0.51$\leq$$x$$\leq$1.0) in the crystals was controlled by the molar ratio 
between LiCoO$_2$ and NO$_2$BF$_4$ ranging from 1 : 0 ($x$=1.0) to 1 : 1.5 ($x$=0.51), and 
was determined by inductively coupled plasma atomic emission spectroscopy (ICP-AES) 
using a Perkin-Elmer Optima 2000 DV instrument. 
For all specimens, a part of the single crystal was crushed into powder and the phase purity of the specimens 
was confirmed by powder x-ray diffraction (XRD) measurements with a 2$\theta$ range from 10$^{\circ}$ to 90$^{\circ}$ 
using a Rigaku RINT2200 diffractometer with Cu $K\alpha$ radiation. 
Almost all of the peaks were found to be ascribed to the Li$_{\rm x}$CoO$_2$ compound. 
XRD patterns of some specimens prepared by the same procedures for 10$^{\circ}$$\leq$$2\theta$$\leq$90$^{\circ}$ 
can be seen in the previous report.\cite{miyoshi10}
dc magnetization measurements were performed by a Quantum Design magnetic property measurement system (MPMS). 
Electrical resistivity was measured by a standard four-probe technique. 
For the resistivity measurements, specimens were cooled by using a Gifford-McMahon (GM) refrigerator or a cryostat of the MPMS. 

\section{Results and Discussion}
\subsection{Powder x-ray diffraction}
The phase mixing of two hexagonal structures 
had been observed earlier in Li$_{\rm x}$CoO$_2$ for 0.75$<$$x$$<$0.90 
through XRD measurements.\cite{menetrier,motohashi09,reimers,ohzuku,amatucci,ishida} 
To investigate the evolution of the phase separation with decreasing $x$, we focus here on the (003) reflection 
observed in the powder XRD measurements. As shown in Fig. 1(a), 
we observe a (003) peak accompanied by a shoulder at the lower angle side for $x$=0.83 and 0.88, 
corresponding to the phase separation, but a (003) single peak for 0.96$\leq$$x$$\leq$1.0. 
Also, we observe a double peak for $x$=0.73 and 0.75 but a single peak again for $x$$\leq$0.64. 
As shown in Fig. 1(b), we can extract two components from the peak for 0.71$\leq$$x$$\leq$0.91. 
We assign the component with a smaller (larger) lattice constant $c$ as the hexagonal phase (I) [hexagonal phase (II)], 
which has higher (lower) Li content.\cite{why1} In Fig. 1(c), the lattice constant $c$ is plotted 
as a function of $x$. Furthermore, we estimated the molar fraction of the hexagonal phase (I) (=$p^{\rm I}$) and (II) (=$p^{\rm II}$) for each $x$ 
from the integrated intensity of the peak. The results are plotted in Fig. 1(d), where it is shown that
the system evolves from phase (I) to phase (II) with decreasing $x$ via the phase mixing state for 0.65$<$$x$$<$0.95.
The variation of $p^{\rm I}$ and $p^{\rm II}$ shown in Fig. 1(d) is consistent with that observed in an earlier work.\cite{ishida}
\begin{figure}[t]
\includegraphics[width=8.5cm]{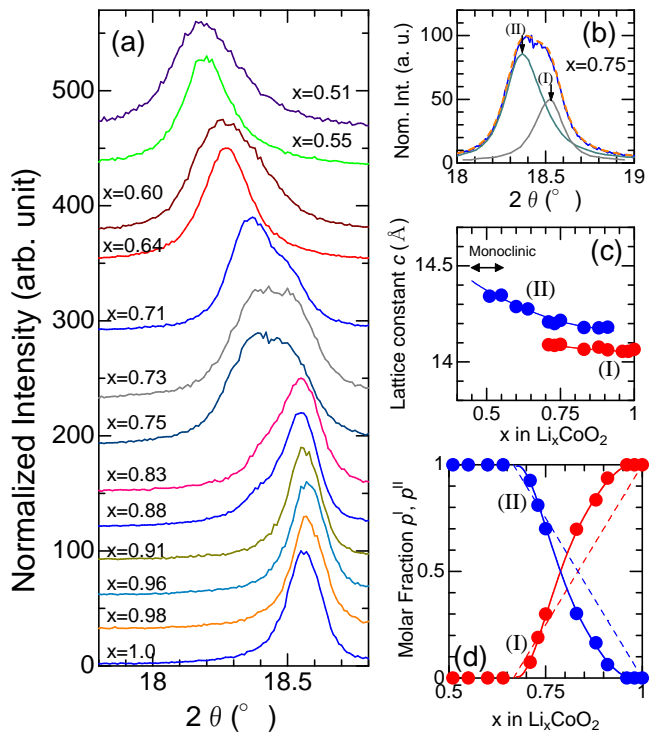}% Here is how to import EPS art
\caption{(Color online) (a) 003 reflection peaks for 
Li$_{\rm x}$CoO$_2$ (0.51$\leq$$x$$\leq$1.0) in powder x-ray diffraction. (b) A typical 003 split peak 
observed due to the phase separation in the specimen with $x$=0.75. 
Two peaks can be extracted from the split peak. 
Each peak corresponds to a hexagonal structure [(I) or (II)] with different lattice parameters. 
(c) Lattice parameters $c$ for Li$_{\rm x}$CoO$_2$ plotted as a function of $x$. The solid lines are guide for the eyes. 
The crystal structure transforms from hexagonal (I) to hexagonal (II) with decreasing $x$. 
A monoclinic phase is known to appear in the neighborhood of $x$=0.5\cite{amatucci}. 
(d) Molar fraction of hexagonal phase (I) and (II) in Li$_{\rm x}$CoO$_2$ estimated from the ratio of 
the integrated intensity of the 003 peak. The solid lines are guide for the eyes. 
The dashed lines indicate the calculated molar fraction of LiCoO$_2$ (red lines) and 
Li$_{2/3}$CoO$_2$ regions (blue line), assuming that the system is composed of the mixture of 
LiCoO$_2$ and Li$_{2/3}$CoO$_2$ for 2$/$3$<$$x$$<$1.0. 
}
\label{autonum}
\end{figure}

Here, we discuss the composition of the hexagonal phases (I) and (II). It is interesting to note that 
a small anomaly at $T_{\rm S}$ which indicates the emergence of the 
CO state is observed even for a slightly delithiated specimen with $x$=0.98 
in the electrical resistivity and dc magnetization data, as seen later in Figs. 2(b) and 3(b). 
These behaviors let us imagine a simple situation that phases (I) and (II) 
are composed of LiCoO$_2$ ($x$=1.0) and Li$_{2/3}$CoO$_2$, respectively, 
and in the latter the CO state emerges below $T_{\rm S}$$\sim$155 K. 
Assuming this, the molar fraction of phases (I) and (II) can be given by the dashed lines shown in Fig. 1(d). 
Although the evolution of the molar fraction with decreasing $x$ is quadratic rather than linear, 
the end of the mixing state appears to be at around $x$=2$/$3 as well as the simple model described by the dotted lines. 
Thus, for 0.9$<$$x$$<$1.0, it is likely that a part of the introduced Co$^{\rm 4+}$ ions by the delithiation 
develops Li$_{2/3}$CoO$_2$ domains [phase (II)], 
while the rest provides holes to the LiCoO$_2$ ($x$=1.0) region [phase (I)], leading to a rapid decrease in 
electrical resistivity with decreasing $x$, as seen later in Figs. 2(a)-2(c). 
The scenario explains why $p^{\rm II}$ increases slower than the dashed line as decreasing $x$. 
As discussed later, phases (I) and (II) can be regarded to have compositions of $x$$\sim$1.0 and $x$$\sim$$2/3$, respectively. 

\subsection{Electrical resistivity}
\begin{figure}[h]
\includegraphics[width=8.5cm]{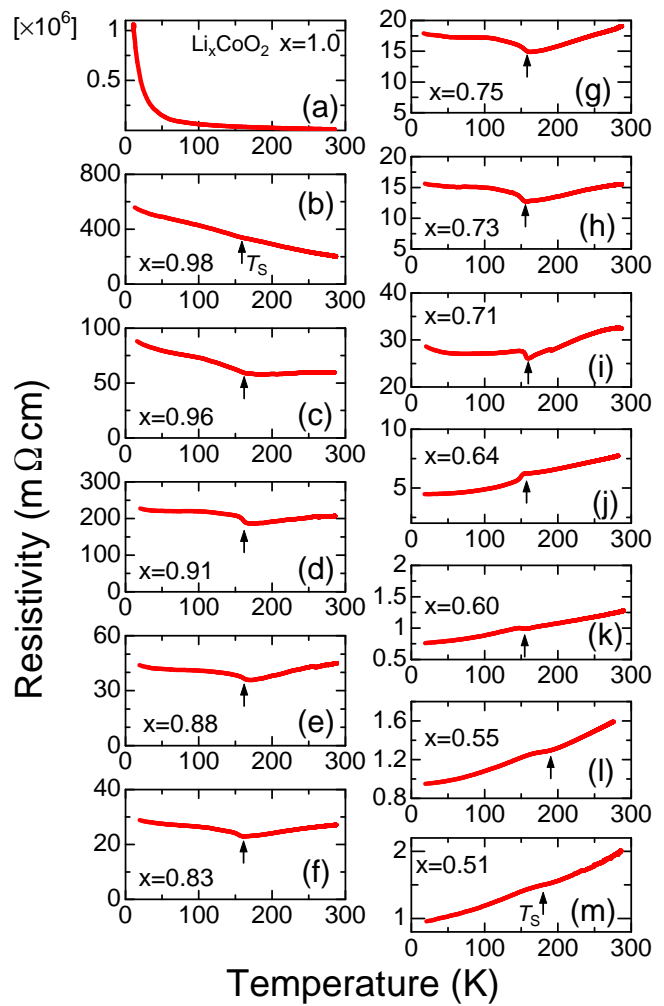}% Here is how to import EPS art
\caption{(Color online) Temperature dependence of electrical resistivity $\rho$ measured 
along the $ab$ plane for Li$_{\rm x}$CoO$_2$ 
with $x$=1.0 (a), 0.98 (b), 0.96 (c), 0.91 (d), 0.88 (e), 0.83 (f), 0.75 (g), 0.73 (h), 
0.71 (i), 0.64 (j), 0.60 (k), 0.55 (l), 0.51 (m). 
}
\label{autonum}
\end{figure}
We show the temperature dependence of electrical resistivity ($\rho$) 
for Li$_{\rm x}$CoO$_2$ (0.51$\leq$$x$$\leq$1.0) in Figs. 2(a)-2(m). The data were collected during heating from $\sim$10 K to 
room temperature after slow cooling of the specimens by using a GM refrigerator at an averaged cooling rate $\sim$$-$2 K$/$min. 
As seen in Fig. 2(a), the $\rho$$-$$T$ curve for LiCoO$_2$ ($x$=1.0) shows an insulating behavior. 
The $\rho$($T$) curves for $x$=0.98 and 0.96 in Figs. 2(b) and 2(c) also show an insulating behavior but the amplitude of resistivity is 
fairly smaller than that for $x$=1.0. We should also note that both of the curves exhibit a slight anomaly at $T_{\rm S}$$\sim$155 K. 
These behaviors are consistent with the picture described in the previous section that the delithiation for 0.9$<$$x$$<$1.0 
gives rise to hole doping in the LiCoO$_2$ region [phase (I)] and also the construction of the Li$_{2/3}$CoO$_2$ domains [phase (II)], 
where the CO state appears below $T_{\rm S}$$\sim$155 K. 

The $\rho$($T$) curves for 0.73$\leq$$x$$\leq$0.91 show a qualitatively similar temperature dependence, 
which is metallic above $T_{\rm S}$ but insulating below $T_{\rm S}$. 
In contrast, the $\rho$($T$) curve for $x$$\leq$0.64 shows a metallic behavior above and below $T_{\rm S}$. 
The metallic behavior below $T_{\rm S}$ is consistent with the results of the DFT calculation and 
the STM observation on the CO state, 
both of which have revealed the metallic electronic structure of the CO state.\cite{iwaya} 
The origin of the insulating behavior observed below $T_{\rm S}$ for $x$$\geq$0.73 is unclear but 
is likely to be related to the fact that the cobalt ions could not build a uniform 
CO state throughout the sample but a disordered one due to an excess of Co$^{\rm 3+}$ for $x$$\geq$2$/$3. 
As mentioned already, an insulating electronic structure has been revealed in the disordered CO state.\cite{iwaya}
It should be also noted that $T_{\rm S}$ changes from 155 K for 0.60$\leq$$x$$\leq$0.98 to 180-190 K for $x$=0.51 and 0.55. 
The different $T_{\rm S}$ values at $x$$\sim$1$/$2 could be a signature of the formation of 
another Co$^{\rm 3+}$$/$Co$^{\rm 4+}$ arrangement with Co$^{\rm 3+}$: Co$^{\rm 4+}$= 1 : 1. 
Hereafter, we call the CO state with a formation of Co$^{\rm 3+}$: Co$^{\rm 4+}$= 2 : 1 and 1 : 1 as 
the CO2$/$3 and CO1$/$2 states, respectively. 
\begin{figure}[h]
\includegraphics[width=7.5cm]{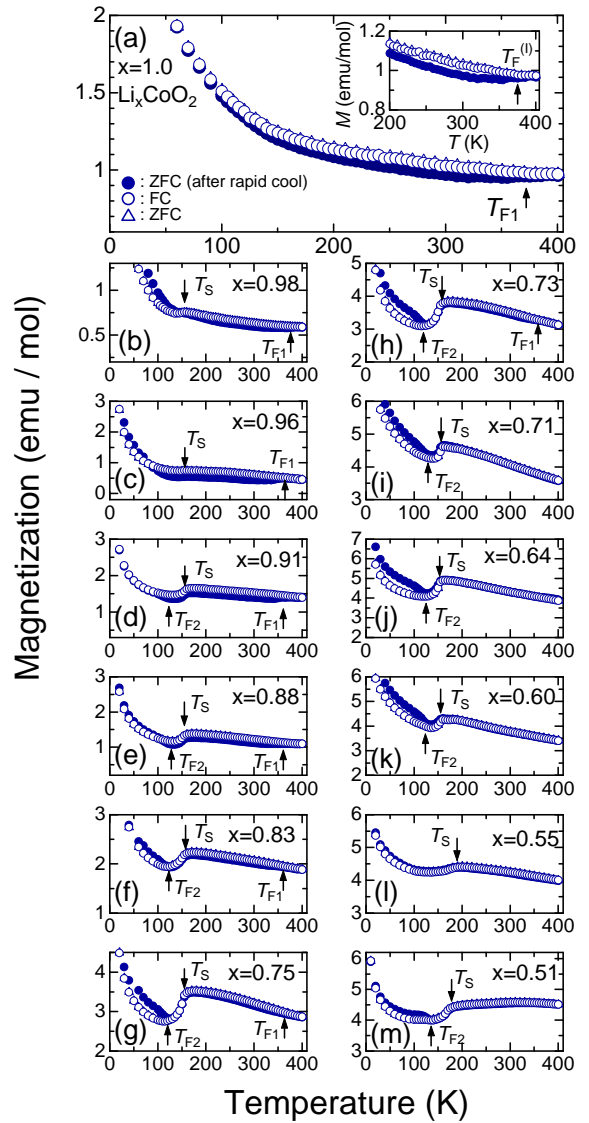}% Here is how to import EPS art
\caption{(Color online) Temperature dependence of dc magnetization 
measured with a magnetic field of $H$=1 T parallel to the $ab$ plane for Li$_{\rm x}$CoO$_2$ with 
$x$=1.0 (a), 0.98 (b), 0.96 (c), 0.91 (d), 0.88 (e), 0.83 (f), 0.75 (g), 0.73 (h), 
0.71 (i), 0.64 (j), 0.60 (k), 0.55 (l), 0.51 (m). The measurements were performed both after rapid cooling of 
the specimens from above 400 to 10 K in zero field (solid symbols) and 
after slow cooling down to 10 K in zero field and in a magnetic field of $H$=1 T (open symbols). 
The inset (a) shows a close-up of the $M$($T$) curve for 200$\leq$$T$$\leq$400 K. 
}
\label{autonum}
\end{figure}

\subsection{dc magnetization}
In Figs. 3(a)-3(m), the temperature dependences of dc magnetization ($M$) measured in a magnetic field parallel to the $ab$ plane under 
zero-field-cooled (ZFC) and field-cooled (FC) conditions for the specimens with $x$=0.51$-$1.0 are shown. 
The data were collected at first with increasing temperature from 10 K 
after rapid cooling of the specimens from 400 to 10 K in zero field, and then collected with increasing temperature after slow cooling 
at a cooling rate of $-$10 K$/$min from 400 K in a magnetic field of $H$=1 T and in zero field. 
For the rapid cooling, the specimens were quickly inserted in the chamber of 
a magnetometer kept at 10 K, after the specimens were taken from an oven heated above 400 K and 
quenched in acetonitrile to room temperature. The $M$($T$) curves for $x$=1.0 in Fig. 3(a) measured after slow cooling 
show a Curie-Weiss type paramagnetic behavior without any difference between the curves in the ZFC and FC conditions.
The $\chi$($T$) [=$M$($T$)$/$$H$] curves are fitted with the formula, $\chi$=$\chi_{0}$ + $C$$/$($T$$-$$\Theta$), 
yielding a constant susceptibility $\chi_{0}$=7.93$\times$10$^{-5}$ emu/mol/Oe, 
a Curie constant $C$=7.29$\times$10$^{-3}$ emu$/$mol$/$K$/$Oe, and a Weiss temperature $\Theta$=$-$4.80 K. 
An effective magnetic moment per Co ions is given to be $\mu_{\rm eff}$=0.241 $\mu_{\rm B}$, which is very small but is similar to 
that obtained in the previous work for $x$=0.99,\cite{miyoshi10} since Co$^{\rm 3+}$ is supposed to be in the low spin state with $S$=0. 

The $M$($T$) curve for $x$=1.0 measured after rapid cooling becomes different from 
those measured after slow cooling below $T_{\rm F1}$$\sim$370 K. 
The $M$($T$) curves for $x$=0.98 and 0.96 in Figs. 3(b) and 3(c) exhibit a slight anomaly at $T_{\rm S}$$\sim$155 K, in addition to the 
difference below $T_{\rm F1}$. Furthermore, the $M$($T$) curves for 0.73$\leq$$x$$\leq$0.91 in Figs. 3(d)-3(h) 
show an abrupt decrease below $T_{\rm S}$, which is followed by an additional splitting 
below $T_{\rm F2}$$\sim$120 K. The $M$($T$) curves for 0.60$\leq$$x$$\leq$0.71 in Figs. 3(i)-3(k) 
are similar to those observed for 0.73$\leq$$x$$\leq$0.91 but have no difference below $T_{\rm F1}$$\sim$370 K. 
As mentioned in the Introduction, $T_{\rm F1}$ and $T_{\rm F2}$ are related to the temperatures, below which Li ions stop diffusing and 
order at the regular site. The difference in the amplitude of $M$ is attributed to the difference in the ordering pattern 
of Co$^{\rm 3+}$$/$Co$^{\rm 4+}$, which depends on 
whether the Li ions sit on the regular site after slow cooling or 
exhibit amorphous-type ordering due to the glass-like freezing of Li$^{+}$ motions after rapid cooling.\cite{miyoshi10,iwaya} 
$T_{\rm F1}$ and $T_{\rm F2}$ seen in Figs. 3(a)-3(k) are thought to be related to the onset of 
the freezing of Li$^{+}$ motions in the 
hexagonal phase (I) with $x$$\sim$1.0 and the phase (II) with $x$$\sim$2$/$3, respectively. 
The $x$ range for which anomalies are observed both at $T_{\rm F1}$ and $T_{\rm F2}$ 
is almost consistent with the nominal phase mixing region (0.7$\lesssim$$x$$\lesssim$0.9).  

It is interesting to note that the amplitude of $M$ measured after rapid cooling is enhanced below $T_{\rm F2}$ but 
coincides with the $M$($T$) curve measured after slow cooling above $T_{\rm F2}$. 
The behavior appears to be remarkable for 0.60$\leq$$x$$\leq$0.83. 
The enhancement of the amplitude of $M$ below $T_{\rm F2}$ is consistent with the destruction of the CO2$/$3 state after 
rapid cooling, since the charge disordered state (liquid state) above $T_{\rm S}$ 
shows an enhanced amplitude compared with that below $T_{\rm S}$. 
One may consider that the intermediate state between $T_{\rm F2}$ and $T_{\rm S}$ after rapid cooling is 
identical to the CO2$/$3 state realized after slow cooling. 
However, it will be seen later in Figs. 4(a) and 4(b) that those states may be similar but are not identical to each other. 
Another interesting feature is that the $M$($T$) curve for $x$=0.55 measured after rapid cooling in Fig. 3(l)
shows a broad anomaly at $T_{\rm S}$$\sim$190 K, much higher than $T_{\rm S}$ of other specimens, 
and no visible anomaly at $T_{\rm F2}$. The $M$($T$) curve for $x$=0.51, however, shows sharp anomalies at 
$T_{\rm S}$$\sim$175 K and $T_{\rm F2}$$\sim$135 K again, both of which are somewhat higher than those for $x$$\geq$0.60. 
As noted in the previous section, the different $T_{\rm S}$ may denote an emergence of the CO1$/$2 state. 
The possible mechanism is discussed later. 
\begin{figure}[t]
\includegraphics[width=7.5cm]{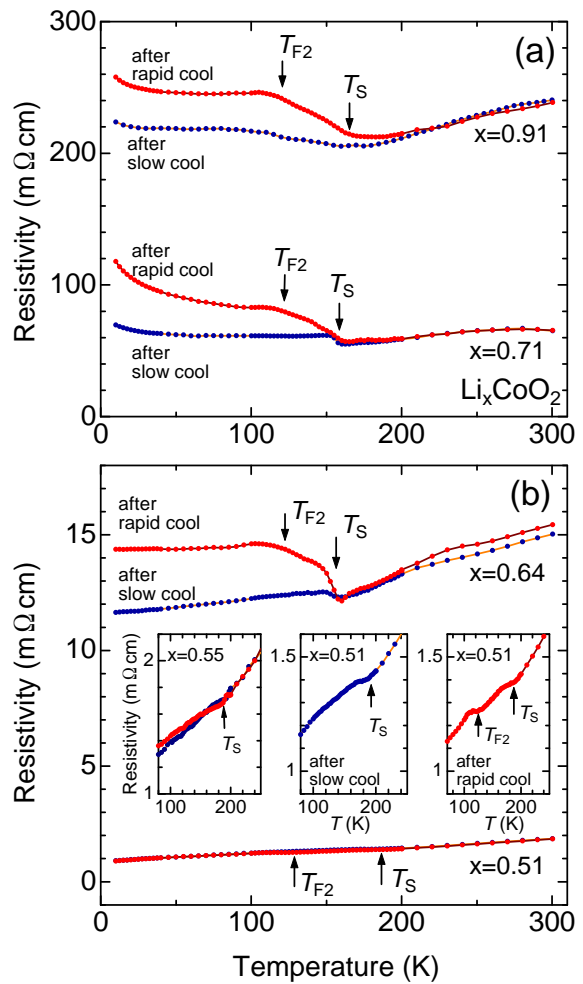}% Here is how to import EPS art
\caption{(Color online) Temperature dependence of electrical resistivity measured 
along the $ab$ plane for Li$_{\rm x}$CoO$_2$ with (a) $x$=0.91 and 0.71 and (b) 0.64 and 0.51. 
The measurements were performed both after slow cooling and rapid cooling of the specimens. 
The insets (b) show close-ups of the $\rho$($T$) curve for $x$=0.51 (right and center) and for $x$=0.55 (left). 
}
\label{autonum}
\end{figure}
\subsection{Enhanced resistivity measured after rapid cooling}
Previous STM observations at $T$$\sim$5 K have revealed an insulating electronic structure 
in the region where disordered arrangements of Co$^{\rm 3+}$$/$Co$^{\rm 4+}$ ions are observed.\cite{iwaya} 
Therefore, in the case where Li ions have an amorphous-like disordered structure after rapid cooling, by which 
the CO$2/3$ state would be destroyed due to the interlayer Coulomb coupling, 
the electrical resistivity is expected to be extremely enhanced compared with 
that after slow cooling. 
Local distortions on the CoO$_2$ layers induced by the amorphous-like structure of the Li ions 
may also contribute to the destruction of the CO2/3 state.
To confirm the behavior described above, 
we have examined the $\rho$($T$) curve after rapid cooling for some specimens. 
For the $\rho$($T$) measurements, we used a cryostat of the MPMS. 
Rapid cooling of the specimens was done by the rapid insertion of the sample rod in the chamber of 
the magnetometer kept at 10 K. 
The results are displayed in Figs. 4(a) and 4(b). 
As seen in the figures, the $\rho$($T$) curves measured after rapid cooling for $x$=0.64, 0.71 and 0.91 
are found to be markedly enhanced compared with those after slow cooling below $T_{\rm S}$. 
The enhancement of resistivity at 10 K attains 70$\%$ of the original value for $x$=0.71. 
Thus, we have certainly confirmed that Li ions probably enter an amorphous-like state after rapid cooling 
and destroy the CO2$/$3 state. A significant difference between the resistivity measured after slow and rapid cooling 
has been observed in some materials that undergo a metal-insulator transition. 
In these materials, a metastable high-temperature low-resistivity state are realized 
at low temperature by rapid cooling after pulsed laser heating.\cite{oike,kagawa} 
If our specimens are cooled more rapidly after local heating by pulsed laser radiation, 
the enhancement is expected to be more significant. 

In contrast, for $x$=0.51, the $\rho$($T$) curve after rapid cooling is not enhanced compared with that after slow cooling, 
as shown in the insets (center and right) of Fig. 4(b). 
For $x$=0.51, an appearance of the CO1$/$2 state is expected but STM observations have never been conducted 
on the specimens with $x$$\sim$0.5, so that the electronic properties of the state are unclear at the present stage. 
The results suggest either that the CO$1/2$ arrangement is not destroyed by the amorphous-like ordering of Li ions or that 
the arrangement is destroyed but the disordered state is also metallic. 
We also show the $\rho$($T$) curves for $x$=0.55 in the left inset of Fig. 4(b), 
where no anomaly is observed at $T_{\rm F2}$ in both curves, which is the same as the $M$($T$) curves for $x$=0.55 in Fig. 3(l). 
We should note here that all the $\rho$($T$) curves after rapid cooling in Figs. 4(a) and 4(b) 
show an anomaly not only at $T_{\rm S}$ but also at $\sim$$T_{\rm F2}$, 
showing a broad peak just below $T_{\rm F2}$. 
On the other hand, $\rho$($T$) curves after slow cooling show an 
anomaly only at $T_{\rm S}$. These features are consistent with those observed in the $M$($T$) curves. 
$T_{\rm F2}$ has been regarded as the temperature below which Li ions stop diffusing and enters into a solid state, 
since a characteristic splitting in the $M$($T$) curves has been observed depending on the cooling process 
below $\sim$120 K.\cite{miyoshi10} However, the $\rho$($T$) curves in Figs. 4(a) and 4(b) 
exhibits a splitting below $T_{\rm S}$, suggesting that Li ions start diffusing above $T_{\rm S}$ rather than $T_{\rm F2}$. 
Considering the interlayer Coulomb coupling, it is more likely that 
the ordering of Li ions also enters a liquid state above $T_{\rm S}$ simultaneously with 
the transition of Co ions from the CO$2/3$ state to a charge liquid state. 
If so, what happens at $T_{\rm F2}$? 
Some amorphous metals, called metallic glasses, demonstrate a two-step transition 
from an amorphous glass state ($T$$<$$T_{\rm g}$) to a liquid state ($T$$>$$T_{\rm x}$) 
via an intermediate supercooled liquid state ($T_{\rm g}$$<$$T$$<$$T_{\rm x}$).\cite{zhang91,peker}  
One of the most exciting scenario is that the intermediate state between $T_{\rm F2}$ and $T_{\rm S}$ 
in Li$_{\rm x}$CoO$_2$ is analogous to a supercooled liquid state observed in metallic glasses. 
To elucidate this, further investigations to confirm the thermodynamical properties by the measurements of specific heat and 
differential scanning calorimetry after rapid cooling of the specimens are desirable for a future study. 
\begin{figure}[t]
\includegraphics[width=8.5cm]{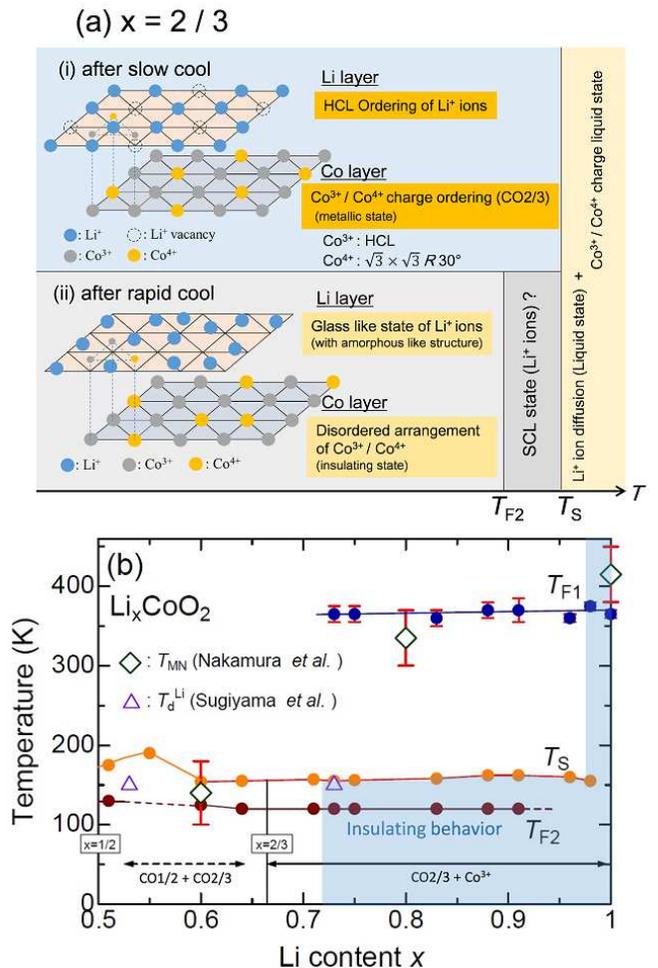}
\caption{(Color online) (a) Schematic image of the ordering in the Li and Co layers 
in Li$_{\rm 2/3}$CoO$_2$ at low temperatures. 
Both Li$^{\rm +}$ and Co$^{\rm 3+}$ ions 
form a honeycomb lattice (HCL) after slow cooling, 
whereas Li$^{\rm +}$ ions enter a glass-like state with an amorphous structure accompanied by the 
disordered Co$^{\rm 3+}$$/$Co$^{\rm 4+}$ arrangement after rapid cooling. The glass-like state is followed 
by a supercooled liquid (SCL) state. 
(b) Plots of $T_{\rm S}$, $T_{\rm F1}$ and $T_{\rm F2}$ vs 
Li content $x$. $T_{\rm MN}$ and $T_{\rm d}^{\rm Li}$, which indicate the onset temperature 
of the Li diffusion determined by the other group\cite{nakamura, sugiyama09}, are also plotted. 
In the blue-shaded area, an insulating behavior is observed in the measurements of electrical 
resistivity. The CO2$/$3 state (Co$^{\rm 3+}$: Co$^{\rm 4+}$= 2 : 1) appears together with an excess of Co$^{\rm 3+}$ ions 
for $2/3$$<$$x$$<$1, while the CO1$/$2 state (Co$^{\rm 3+}$: Co$^{\rm 4+}$= 1 : 1) appears at $x$$\sim$1$/$2. 
These CO states are thought to coexist for a certain $x$ range around $x$$\sim$0.55. 
}
\label{autonum}
\end{figure}

\subsection{Ordering features and phase diagram}
Here, we show the schematic image of the ordering in the Li and Co layers for $x$=2$/$3 at low temperatures in Fig. 5(a). 
Above $T_{\rm S}$, Li ions can diffuse in the layers, and Co$^{\rm 3+}$$/$Co$^{\rm 4+}$ ions are in a charge liquid state. 
After slow cooling, Li$^{\rm +}$ and Co$^{\rm 3+}$ ions form a honey comb lattice (HCL) below $T_{\rm S}$, while Co$^{\rm 4+}$ ions 
form a characteristic $\sqrt{3}$$\times$$\sqrt{3}$$R$30$^{\circ}$ structure.\cite{iwaya} 
Contrasting to the regular structures after slow cooling, Co$^{\rm 3+}$$/$Co$^{\rm 4+}$ ions have 
a disordered arrangement after rapid cooling, as discussed in the previous section. 
Also, it is inferred that Li ions are frozen in an amorphous state from a liquid state by rapid cooling, 
leading to a disordered Co$^{\rm 3+}$$/$Co$^{\rm 4+}$ arrangement due to the interlayer Coulomb coupling. 
Amorphous structures in the Li layers have been also inferred for the specimen with $x$=0.99 
to explain the large thermal history dependence in the $M$($T$) curve below $\sim$380 K, 
which originates from the difference in the ordering pattern of Li ions depending on the cooling rate.\cite{miyoshi10} 
In the case that the Li site is fully occupied, the only way to give a different Li ordering pattern is to build an 
amorphous structure. For $T_{\rm F2}$$\leq$$T$$\leq$$T_{\rm S}$, 
we propose that the Li ions enter a supercooled liquid (SCL) state. 

Next, we focus on the $x$ dependences of $T_{\rm F1}$, $T_{\rm F2}$ and $T_{\rm S}$, which are summarized in Fig. 5(b). 
We notice that these temperatures are almost independent of $x$, suggesting that the hexagonal phase (I) 
[phase (II)] has a similar composition of LiCoO$_2$ (Li$_{2/3}$CoO$_2$) 
throughout for 0.73$\leq$$x$$\leq$1.0 (0.6$\leq$$x$$\leq$0.98). 
In Fig. 5(b), we also plot $T_{\rm MN}$ and $T_{\rm d}^{\rm Li}$ determined by $^7$Li
NMR\cite{nakamura} and $\mu$$^+$SR experiments\cite{sugiyama09}, respectively, above which Li ions start diffusing. 
We note here that $T_{\rm MN}$ for $x$=0.8 has been observed only for hexagonal phase (I),\cite{nakamura} so that 
$T_{\rm MN}$ has a single value even in the two-phase region. 
$T_{\rm MN}$ and $T_{\rm d}^{\rm Li}$ are shown to be consistent with $T_{\rm F1}$ and $T_{\rm F2}$. 
Finally, we discuss how the charge ordering changes with decreasing $x$ from $x$=$2/3$, 
since an appearance of the CO$1/2$ state is expected at $x$$\sim$0.5. 
The important point to note is the anomalous behaviors at $x$=0.55, where $T_{\rm S}$ is fairly enhanced 
and the anomaly at $T_{\rm F2}$ is absent in the $\rho$($T$) and $M$($T$) curves. 
The most plausible scenario is that the CO$2/3$ state survives with an excess of Co$^{\rm 4+}$ for $x$$<$$2/3$, 
but the CO$1/2$ state begins to coexist below certain $x$ and competes with the CO$2/3$ state near $x$=0.55, 
and then the CO$1/2$ state becomes dominant for $x$$\sim$$1/2$. 
Due to the competition between the CO$2/3$ and CO$1/2$ states, both arrangements of 
Li$^{\rm +}$ and Co$^{\rm 3+}$$/$Co$^{\rm 4+}$ could be disordered even after slow cooling.  
Under the circumstance, we may expect either that Li ions do not undergo the transition at $T_{\rm F2}$ or 
that the arrangement of Co$^{\rm 3+}$$/$Co$^{\rm 4+}$ remains disordered even across $T_{\rm F2}$. 
In both cases, no anomaly would be observed at $T_{\rm F2}$. 
Assuming that all Co$^{\rm 3+}$$/$Co$^{\rm 4+}$ ions contribute to either CO$1/2$ or CO$2/3$ domains, 
the critical value of $x$ at which the occupied area by the CO$1/2$ and CO$2/3$ states is totally equivalent 
(i.e., the molar fraction $p^{\rm CO1/2}$ $:$ $p^{\rm CO2/3}$ = 3 $:$ 2) is estimated to be 7$/$12$\sim$0.583. 
The value is close to 0.55. 
The elucidation for the CO$1/2$ state and the ordering of Li ions at $x$=0.5 by the combined study of 
the STM observation and DFT calculation must be the important step to understand the overall feature of Li$_{\rm x}$CoO$_2$

\section{Summary}
In the present paper, we have investigated the low temperature magnetic and electrical properties of 
Li$_{\rm x}$CoO$_2$ (0.51$\leq$$x$$\leq$1.0) using single-crystal specimens.  
The CO$2/3$ state has been found to appear in the wide $x$ range 0.60$\leq$$x$$\leq$0.98. 
In the powder XRD measurements, we have observed mixed hexagonal phases (I) and (II) for 0.71$\leq$$x$$\leq$0.91, 
each of which has a different lattice parameter $c$, in other words, different Li content. 
Taking also the results of the $\rho$($T$) and $M$($T$) measurements into account, it is suggested that 
the composition of phase (I) [phase (II)] is nearly LiCoO$_2$ (Li$_{\rm 2/3}$CoO$_2$), 
where the Li ions start diffusing above $T_{\rm F1}$$\sim$370 K ($T_{\rm F2}$$\sim$120 K).
Furthermore, we have successfully observed that the $\rho$($T$) curves are fairly enhanced 
after rapid cooling of the specimens below $T_{\rm S}$$\sim$155 K, as an evidence of the occurrence of the disordered 
Co$^{\rm 3+}$$/$Co$^{\rm 4+}$ arrangement with an insulating electronic structure, which is led by an amorphous-like formation of 
Li ions. It is inferred that the CO$1/2$ state, which has an arrangement with 
Co$^{\rm 3+}$$:$ Co$^{\rm 4+}$=1 $:$ 1, is realized at $x$$\sim$$1/2$ 
and competes with the CO$2/3$ state for a certain $x$ range around $x$=0.55.   
We have also suggested that the state after rapid cooling between $T_{\rm F2}$ and $T_{\rm S}$ can be regarded as a 
supercooled liquid state as seen in metallic glass materials. 
Finally, we should note that Li$_{\rm x}$CoO$_2$ is a high energy cathode material for Li ion batteries, but also 
an electron system, which provides us a great opportunity to encounter unique low-temperature 
properties related to the Li ion dynamics, being a potential material to exhibit 
a rapid-cooling-induced giant resistivity which is led by a different mechanism. 
Great attention should be paid to the progress of these studies. 

\begin{acknowledgments}
The authors thank H. Katsube and K. Mihara for their technical assistance.  
\end{acknowledgments}

%\begin{thebibliography}{99}


\begin{thebibliography}{99}
\bibitem{why1}
Generally, layered oxides $A$$_{\rm x}$CoO$_2$ ($A$=alkali metal) have a tendency to have a larger lattice constant $c$ for smaller $x$, 
because the interlayer Coulomb interactions are reduced and the interlayer distance is increased with decreasing the concentration of 
alkali ions.
\end{thebibliography}
\end{document}